\documentclass[11pt]{article}
\usepackage[dvips]{graphics}
\begin{document}

\newcommand{\be}{\begin{equation}}
\newcommand{\ee}{\end{equation}}
\newcommand{\bea}{\begin{eqnarray}}
\newcommand{\eea}{\end{eqnarray}}
\newcommand{\Tr}{\mbox{Tr}}
\newcommand{\bc}{\begin{center}}
\newcommand{\ec}{\end{center}}
\newcommand{\R}{\vec{R}}
\newcommand{\x}{\vec{x}}

\title{String-breaking in quenched QCD}

\author{Chris Stewart and Roman Koniuk \\ 
Physics and Astronomy,
        York University, Toronto, Canada.}
       
\maketitle

\begin{abstract}
We present results on a new operator 
for the investigation of string-breaking effects in quenched SU(2)-colour QCD.
 The ground-state 
of a spatially-separated static-light meson-antimeson pair is a combination 
of a state with two distinct mesons, expected to dominate for large 
separations, and a state where the light-quarks have annihilated, which 
contributes at short distances. The crossover between these two regimes 
provides the string-breaking scale.
\end{abstract}

\section{Introdction}

An early success of lattice QCD simulations was the first-principles 
demonstration that the potential between static quarks is confining.
The binding energy in the static quark-pair system is described very well 
by a coulomb-plus-linear potential. 
Recent studies have confirmed the expectation that the gauge field forms a 
narrow flux tube, or string, joining the static quarks \cite{Pennanen}.

In unquenched simulations, we expect the static-quark potential
to show evidence for string-breaking---when the energy in the gauge field 
string allows the creation of a light quark pair from the vacuum, 
the system should `break' into a static-light meson-antimeson pair. The signal
of string-breaking would be a plateau in the static-quark potential, around the
mass of the static-light meson.

A first-principles demonstration of string-breaking has so far eluded 
lattice QCD practitioners. Traditionally, researchers have performed Wilson 
loop simulations in
full QCD, searching for the signature plateau in the static quark potential.
Recent results show no evidence for this effect \cite{Gusken}.
This has led some to
suggest that the Wilson loop operator may have too small an overlap with the
broken two-meson state \cite{Gusken,de Forcrand}, and to recommend a 
search for better operators.

In a previous paper, we described a derivation of the binding potential 
between two static-light mesons in the quenched approximation of SU(2) QCD
\cite{hadmol}. A similar investigation of the binding potential in a static
meson-antimeson pair was abandoned due to difficulties with the operator's
short-range
behaviour---the derived binding potential appeared identical to the Wilson
loop potential, which we interpreted as a signal that the light quarks were
annihilating, leaving only a static quark-pair.

The light-quark annihilation led us to conclude that the meson-antimeson 
operator, while perhaps
unsuitable for an investiagtion of the binding potential, might be ideal for a 
demonstration of string-breaking phenomena. This approach is a reversal of the
usual Wilson loop approach---we know that our operator has the correct
long-distance behaviour, and wish to show it also describes the 
unbroken-string short-distance behaviour of the Wilson loop.

Our aim in this paper is to demonstrate that the static-light  
meson-antimeson operator is well-suited to string-breaking applications, 
providing 
superior overlap with the broken-string state while retaining the necessary 
overlap with the unbroken quark-pair state. We describe the operator and the 
simulation parameters in the following sections, and then present preliminary 
results that indicate the operator does describe 
string-breaking within the quenched approximation. This result makes
us confident that the operator will be useful in full-QCD simulations.

\section{The Operator}

The standard operator used in string-breaking investigations is the 
Wilson loop, which for a closed loop $C(R,T)$ of spatial width $R$ and time
extent $T$ is
\be
W(R,T) = \Tr \prod_{l \in C_{R,T}} U_{l} \, .
\ee
The Wilson loop is the propagator for a spatially-separated static 
quark-antiquark pair.
By design, this operator has strong overlap with the unbroken state of
two static quarks joined by a gluon flux tube---sadly, it has proven to have
insufficient overlap with the broken state of two distinct static-light mesons,
as recently demonstrated by Knechtli and Sommer \cite{Knechtli}. 
The lack of evidence for
string-breaking is most likely due to this poor overlap
with the broken state: the meson-pair state simply isn't `seen' by the 
Wilson loop operator.

Consider then, a composite operator consisting of a static-light 
meson-antimeson pair, separated by a distance $\R$,
\be
\label{op}
{\cal O}(\R) = \bar{\psi}_l (0) \Gamma \psi_S (0) 
\bar{\psi}_S (\R) \Gamma \psi_l (\R) \,.
\ee
We used a smeared-source meson operator, with
\be
\label{smear}
\Gamma = \gamma_5 (1+\epsilon\Delta^2)^{n_s}
\ee
to improve the overlap with the meson ground-state \cite{Howard}.
The meson-pair correlator is
\be
\label{mmbarp}
G(t,\R) = {\cal G}_{D} + {\cal G}_{E} \, ,
\ee 
where
\bea
{\cal G}_{D}(t,\R) &=& \Tr \left [ G_{h}(0,t;0,0) \, G^{\dag}_{l}(0,t;0,0) 
\right ] \nonumber \\
& \times &  \Tr \left [ G_{l}(\R,t;\R,0) \, G^{\dag}_{h}(\R,t;\R,0) \right ] 
\nonumber \,, \\
{\cal G}_{E}(t,\R) &=& - \Tr \left [ G_{h}(0,t;0,0) \, 
G^{\dag}_{l}(\R,t;0,0) \right .
\nonumber \\
& & \left .G^{\dag}_{h}(\R,t;\R,0) \, G_{l}(0,t;\R,0) \right ] \, .
\eea
Contributions to ${\cal G}_{D}$ and ${\cal G}_{E}$, the `direct' and 
`exchange' terms, are depicted in Figure 1. 

In the large-$R$ limit, the operator describes two distinct static-light 
mesons, and so we expect ${\cal G}_D$ will dominate the correlator for large 
separations. For small 
$R$, however, the light quarks can easily annihilate, as shown in Figure 2,
leaving a static quark-antiquark pair interacting through the gluon field.
The exchange term ${\cal G}_E$ should contribute strongly for small 
separations. 
The operator in equation (\ref{op}) provides a description of the 
long-distance physics 
of the broken mesonic state, {\em and} the short-distace physics of the 
static-quark pair thanks to the annihilation of the light quarks. 

\begin{figure}[htb]
\begin{center}
\vspace{9pt}
\scalebox{0.6}[0.6]{\includegraphics{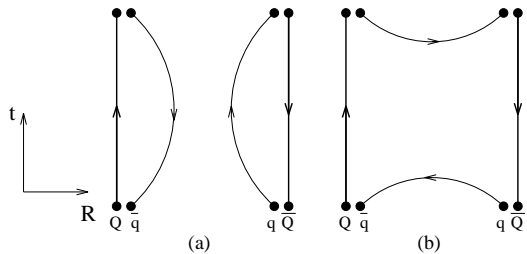}}
\caption{Contributions to (a) direct and (b) exchange terms.}
\end{center}
\end{figure}

\begin{figure}[htb]
\begin{center}
\vspace{9pt}
\scalebox{0.6}[0.6]{\includegraphics{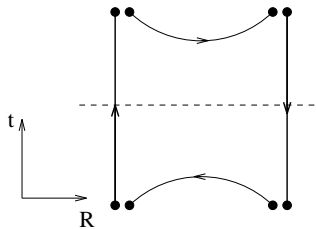}}
\caption{Light-quark annihilation in the exchange term---at the time slice
indicated by the dashed line, only the static quarks remain.}
\end{center}
\end{figure}

\section{The Simulation}

We performed a lattice simulation to test the validity of equation 
(\ref{op}) 
as a string-breaking operator. 
An ensemble of 344 quenched SU(2)-colour gauge configurations was
created using an ${\cal O}(a^2)$-improved action,
\be
S_{G} = -\beta \sum_{x,\mu > \nu} \left (\frac{5}{3}\frac{P_{\mu\nu}}
{u_0^4} - \frac{1}{12} \frac{R_{\mu\nu} + R_{\nu\mu}}{u_0^6} \right ) \, .
\ee
where $P_{\mu\nu}$ is the plaquette operator, and $R_{\mu\nu}$ is a $2\times 1$
loop with the long side along the $\mu$-direction. The lattice dimensions were
$(L_x,L_y,L_z,L_t) = (10,8,8,12)$, with the separation between the static
quarks along the x-direction. 
The tadpole correction $u_0$ was derived from the plaquette operator,
\be
u_0 = \langle P_{\mu\nu} \rangle^{1/4}\, .
\ee
The simulation was performed at
$\beta = 1.07$, corresponding to a lattice spacing of roughly $0.2 fm$, 
using the $\rho$- and $\pi$-meson mass ratio to set the scale.

We used the tadpole-improved Sheikholeslami-Wohlert operator for the
fermion action,
\be
M_{SW} = m_0 + \sum_{\mu} \left (\gamma_{\mu} \triangle_{\mu} - 
\frac{1}{2}
\triangle^2_{\mu}\right ) - \frac{1}{4} \sigma \cdot F \, .
\ee
The correlator, equation (4), requires us to determine light-quark propagators
for each value of R and T---this computational complexity forced us to choose 
a high mass for the light-quarks. We used $\kappa = 0.135$, corresponding to a 
pion to rho-meson mass ratio of $m_{\pi}/m_{\rho} \simeq 0.76$. 
We chose $\epsilon = 1/12$ and $n_s = 5$ in the smearing function, 
eq. (\ref{smear}) \cite{Howard}. 

\section{Results}

All masses and energies were taken from single-cosh fits to the 
relevant propagators. Results from multi-cosh fits were not reliable, 
though this is hardly surprising given the relatively small number of 
propagator elements, due to the small dimensions of the lattice.

Figure 3 shows a comparison of the direct and exchange
terms in the propagator (\ref{mmbarp}) for varying separation $R$. Also shown
is  a linear-plus-coulomb fit to the Wilson loop data from the same lattice
ensemble. 
Note that for small
R, the exchange term gives a contribution almost identical in size to the 
Wilson loop potential, indicating that the light quarks are indeed annihilating
to leave a static quark-antiquark pair.

\begin{figure}[tb]
\begin{center}
\vspace{9pt}
\scalebox{0.33}[0.33]{\includegraphics{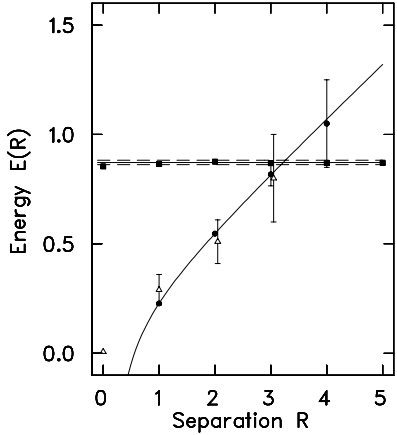}}
\caption{Direct term (squares),
exchange term (triangles) and Wilson loop potential 
(circles and solid line). Exchange term data are 
offset for clarity. Horizontal lines indicate static-light meson mass.}
\end{center}
\end{figure}

We expect string-breaking to occur at the point where the two-meson state 
becomes  energetically
favourable---from the slope of the Wilson loop potential,
this appears to be between $R = 3$ and $R = 4$.
Unfortunately, noise overcomes the exchange-term signal just at this point, 
and the crossover can only be inferred from the Wilson loop data. 

The full propagator, shown in Figure 4, provides a
much clearer view of the potential. The system's energy increases to a 
plateau at the
expected level of the mass of the two distinct mesons, indicated by the 
horizontal lines. The static-light meson mass was taken from fits to the 
single-meson propagator.
The small error bars on the potential
for large values of $R$ indicate that, although noise has destroyed the 
exchange term's signal, the mixing into this term has vanished. This is again 
expected, since the meson-pair state should dominate as R increases.

From Figures 3 and 4, we estimate the string-breaking distance to be
roughly $3a$, or $0.6$ fermi, using our naive SU(2) lattice spacing estimate.

\begin{figure}[tb]
\begin{center}
\vspace{9pt}
\scalebox{0.33}[0.33]{\includegraphics{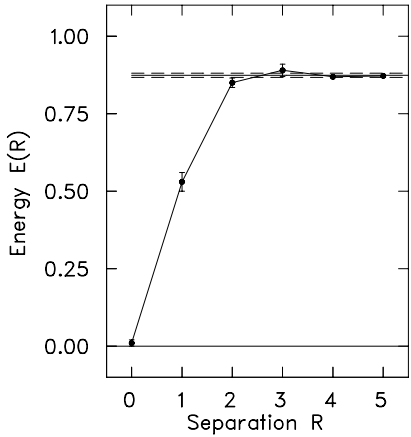}}
\caption{Potential from full propagator. Horizontal lines indicate mass and 
error limits of the free static-light meson.}
\end{center}
\end{figure}

\begin{figure}[tb]
\begin{center}
\vspace{9pt}
\scalebox{0.33}[0.33]{\includegraphics{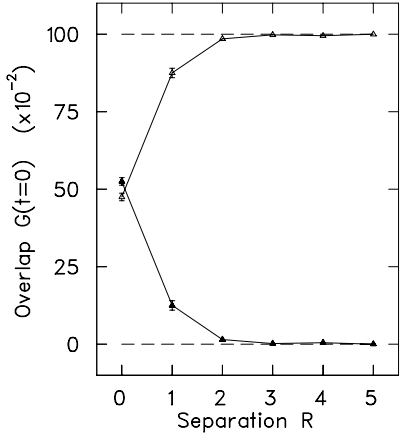}}
\caption{${\cal G}_D(t = 0)$ and ${\cal G}_E(t=0)$ as a measure of the overlap 
with the ground state.}
\end{center}
\end{figure}

Figure 5 supports this interpretation. The comparative sizes of 
${\cal G}_D(t = 0)$ and
${\cal G}_E(t = 0)$ are taken as a rough measure of the mixing
of each of the
direct and exchange terms with the ground state, and are plotted as a function
of $R$. As expected,
the direct and exchange terms both contribute for small $R$, 
and the exchange term vanishes quickly as separation
increases, leaving the direct term to dominate completely for large $R$.

\section{Conclusions}

The use of new operators, or combinations of operators, to demonstrate
string-breaking on the lattice appears to be an idea whose time has 
arrived---witness the flood of recent papers presenting results
in $SU(2)$ QCD with scalar fields \cite{Knechtli,Trottier,Philipsen}, 
and references to ongoing research in full $SU(3)$ QCD \cite{Pennanen}. 

We have described an operator suitable for use in string-breaking 
investigations. The operator, equation (\ref{op}), 
is able to describe both regimes necessary for a demonstration of 
string-breaking---the short-range static-quark pair, and the 
long-range meson pair. 
For small separations, this operator behaves like a Wilson loop, 
thanks to the annihilation of the light quarks, resulting in 
a confining potential. For larger separations, the potential reaches a 
plateau at the energy of two distinct static-light mesons, the 
fingerprint of the elusive broken string. 
Our simulation gives a string-breaking distance of $\sim 0.6 fm$, and noting
that we are working with SU(2)-colour gauge fields, this result is in 
qualitative agreement with other estimates \cite{Knechtli}.

The string-breaking occurs even in our {\em quenched} simulation, since the
operator {\em must} energetically favour the meson-pair state for large $R$, 
and so forces the gluon string to break. The light quarks are
providing, within the quenched approximation, some sea-quark effects. We
expect the operator will provide the same results in unquenched simulations.

Our computing resources forced certain constraints on our 
simulations---performing this research on a desktop workstation, we chose 
a high light-quark mass, small lattice volume and large lattice spacing. 
To counter the effects of low statistics and finite lattice spacing, we 
employed improved actions and operators. A truly definitive demonstration 
of string-breaking would require much higher statistics and finer resolution 
than the results presented here.

While we are acutely aware of the
limitations of the results described here, we reiterate our main goal---to
demonstrate the
utility of the static meson-antimeson operator in string-breaking simulations. 
We are confident that this operator
will allow accurate determination of the string-breaking distance when used
in more ambitious simulations.

We thank Howard Trottier and Norm Shakespeare for helpful discussions and 
suggestions. This work was supported in part by the National Sciences and 
Engineering Research Council of Canada.


\begin{thebibliography}{9}
\bibitem{Pennanen} P. Pennanen, to appear in proceedings
of Lattice 98, Boulder Colorado (1998), and hep-lat/9809035.
\bibitem{Gusken} Stephan G\"{u}sken, Nuc. Phys. B (Proc. Suppl.) 63 (1998) 16.
\bibitem{de Forcrand} Ph. de Forcrand, to appear
in proceedings of Lattice 98, Boulder Colorado (1998).
\bibitem{hadmol} Chris Stewart and Roman Koniuk, Phys. Rev. D 57 (1998) 5581.
\bibitem{Knechtli} Francesco Knechtli and Rainer Sommer, to appear in 
proceedings of Lattice 98, Boulder Colorado (1998), and hep-lat/9807022.
\bibitem{Howard} Norman Shakespeare and Howard Trottier, Phys. Rev. D 58 
(1998) 034502, and hep-lat/9802038.
\bibitem{Trottier} Howard Trottier, to appear in proceedings
of Lattice 98, Boulder Colorado (1998), and hep-lat/9809183. 
\bibitem{Philipsen} Owe Philipsen and Hartmut Wittig, to appear in proceedings
of Lattice 98, Boulder Colorado (1998), and hep-lap/9807020.
\end{thebibliography}
\end{document}